\documentclass[preprint,showpacs,preprintnumbers,amsmath,ammsymb]{revtex4}

\draft

\begin{document}

\title{
Tuning electronic structure of graphene via tailoring structure: theoretical study
}
\author{
         $H.Y. He^1$, $Y. Zhang^1$ and $B.C. Pan^{1,2}$
}
\affiliation{
  1 Department of Physics,
  2 Hefei National Laboratory for Physical Sciences at Microscale,  \\
   University of Science and Technology of China,\\
 Hefei, Anhui 230026, People's Republic of China \\
}


\begin{abstract}

Electronic structures of graphene sheet with different defective patterns are investigated, 
based on the first principles calculations. We find that defective patterns can tune the 
electronic structures of the graphene significantly. Triangle patterns give rise to strongly localized 
states near the Fermi level, and hexagonal patterns open up band gaps in the systems. 
In addition, rectangular patterns, which feature networks of graphene nanoribbons with either zigzag 
or armchair edges, exhibit semiconducting behaviors, where the band gap has an evident dependence on the 
width of the nanoribbons. For the networks of the graphene nanoribbons, some special channels 
for electronic transport are predicted.
 
\end{abstract}


\maketitle

\section {INTRODUCTION}
Graphene has been a hot issue in both fundamental and application research fields, since stable 
single graphene layer at room temperature was achieved 
experimentally \cite{Novoselov-1,Novoselov-2,Zhang,Stankovich}. Such two-dimensional hexagonal honeycomb has 
novel electronic \cite{Geim}, mechanical \cite{Lee} and thermal\cite{Balandin} properties. 
For example, a perfect graphene sheet is semimetallic, being an extremely good conductor. 
Meanwhile, graphene exhibits astonishing transport properties: electron mobility 
in graphene was predicted to be as high as $15000$ $cm^2/Vs$ at room temperature. 
Moreover, atomic thickness, chemical inertness and almost perfect structure makes it promising material for device 
application. Therefore, graphene holds a sound prospects for a new generation of graphene-based 
electronics \cite{Rycerz,Ponomarenko,Wang,Lin,Avouris}.

However, most electronic and optoelectronic devices require a semiconductor with a finite gap.
To broaden its applications, many efforts have been devoted to turn the semimetallic graphene 
into a gapped semiconductor, through loading strain \cite{Pereira}, introducing defects \cite{Topsakal}, and 
performing chemical functionalization \cite{Wu,Yan}. 
Of these efforts, designing superlattice scale defects in graphene sheet, so called
nanopattering, has drawn much attention.

Experimentally, patterning of a graphene sheet by e-beam lithography with features as small as ten nm 
was available \cite{Elias,Berger,Han}, indicating that patterning graphene with the desired features is possible. 
In fact, a quasi-one-dimensional graphene nanoribbon (GNR) has been successfully achieved, and its 
electronic structures have a strong dependence on the width and chirality \cite{Abanin,Okada,Louie}.
Quite recently, nanoscale square array of holes have been fabricated on the graphene in experiment, and 
the existence of a transport gap was found \cite{Shen,Eroms}.

On the theoretical side, graphene antidot lattices, graphene sheets with regularly spaced holes, 
have been proposed as a platform for quantum confinement of electrons in graphene \cite{Pedersen-1,Pedersen-2,Vanevic}.  
Based on the tight-binding (TB) calculations, the electronic structures of the antidot lattice with triangular arrays 
 or roughly holes in graphene sheet were illustrated by two groups \cite{Pedersen-1,Vanevic}. 
Such arrays of nanoscale perforations in graphene manipulate the electronic structures of the systems significantly, 
by giving rise to the localized states, or inducing the band gap near the Fermi level. Very recently, such antidot lattices 
proposed by Pedersen $et$ $al$ were more accurately studied by using the density functional theory 
calculations \cite{Furst}. Band gaps ranging from 0.2 to 1.5 eV were reported.
However, all these concerned antidots were arranged in a hexagonal unit cell. If the triangular (hexagonal) 
holes are arranged in a rectangular supercell, the antidots display different geometry features. 
Whether do the array features have significant influence on the electronic behaviors ?

On the other hand, although square arrays have been achieved experimentally, theoretical efforts 
on such defective patterns have not been reported yet, to our best knowledge.
Structurally, such square arrays behave as the networks of armchair edged GNRs (a-GNRs) and zigzag edged GNRs (z-GNRs).
As we know, a-GNRs and z-GNRs display different electronic properties: the former exhibits
an interesting three-family behavior, while the latter has a direct band gap which decreases
with increasing width of the z-GNR \cite{Louie}. For the networks consisting of a-GNRs and z-GNRs, 
electronic properties of the systems are still puzzling.

In this work, we firstly investigate the electronic structures of graphene sheet with triangular and hexagonal holes 
in a rectangular supercell, based on the first principles calculations. Then rectangular holes are introduced 
into a graphene sheet, which feature the networks of a-GNRs and z-GNRs. We systematically explore 
the electronic structures of such patterned graphene with varying 
the width of either a-GNR or z-GNR. It is found that the graphene sheets with rectangular holes are all 
semiconductors. Meanwhile, tuning band gap can be practiced through different patterning.
Moreover, the size-effect of different structural feature on electronic structures is also discussed.

\section {COMPUTATIONAL METHOD}

Our calculations are performed by using the SIESTA program \cite{Siesta} at the level of local density approximation,
in which the norm-conserving pseudopotential are taken into account. Spin-polarized calculations are performed 
for all our concerned cases.
For both $C$ and $H$ atoms, double- $\zeta$ basis sets \cite{Soler} are chosen for calculations. 
For each concerned system, the periodic boundary conditions along $x$ and $y$ axes are considered. Meanwhile, 
a spacing of more than 10 $\AA$ is applied along $z$ axis to neglect the interaction between 
the graphene monolayers.
All atomic positions and the lattice constants of the concerned systems are allowed to be fully optimized, 
with the residual force convergence value of less than 0.02 eV/$\AA$.
The Brillouin zone is sampled with $12\times12\times1$ according to the Monkhorst-Pack scheme \cite{Monkhorst}, 
which is tested to be enough for our calculations. In the investigations of the electronic structures, 
more k-points with sampling of $18\times18\times1$ are employed.

\section {ELECTRONIC STRUCTURES}

Three kinds of structural patterns (triangular, hexagonal and rectangular holes) in graphene sheets are concerned, 
which are shown in Fig.~\ref{Fig.1}. For the systems with triangular or hexagonal holes, the lattice constants 
of the supercell are chosen to be 25.56 and 24.60 $\AA$ along $x$ axis and $y$ axis respectively.
 While for the systems with rectangular holes, they are set to be 25.56 and 19.68 $\AA$.
Fig.~\ref{Fig.1}-(a) is a generated graphene with a triangular void by removing a 
triangular portion with zigzag edges.
In addition to the graphene with absence of a triangular portion shown in Fig.~\ref{Fig.1}-(a), three more cases 
with different size of the voids are considered, which are formed by removing the carbon atoms 
at the edges marked with $E1$, $E2$ and $E3$ in Fig.~\ref{Fig.1}-(a) in turn. 
For convenience, we notate the systems with triangular holes from small size to large size as tri-1, tri-2, tri-3 
and tri-4 respectively. In our calculations, the carbon atoms with two-folded coordinates near the edges are terminated with 
hydrogen atoms for all our concerned cases.

The calculated density of states (DOS) for the cases of tri-1, tri-2, tri-3 and tri-4 are 
illustrated in Fig.~\ref{Fig.2}-(a-d), in which the DOS of the corresponding perfect sheet is plotted as 
a reference (in Fig.~\ref{Fig.2}-(d)). Clearly, the majority spin states and the minority spin states 
split, and the peaks strikingly appear near the Fermi level, which is consistent with the previous report \cite{Furst}. 
For the cases of tri-1 and tri-2, two evident peaks emerge at about $E_{f}-0.1 eV$ and $E_{f}+0.1 eV$ in 
the calculated DOS respectively. With increasing the size
of the defects, the peak under the Fermi level becomes broad and evolutes to be two peaks eventually. This is 
shown obviously in Fig.~\ref{Fig.2}-(d) for the case of tri-4. 
The further analysis reveals that these peaks near the Fermi level are mainly contributed from the carbon 
atoms in the outmost row at the defect edges which bond to hydrogen. For convenience, we refer these atoms 
as the edge atoms. Moreover, we find that most carbon atoms which bond with three other carbon atoms
contribute little to these peaks, whether these three coordinated carbon atoms are close to or far away from the edges.

Comparing the DOS of tri-1, tri-2, tri-3 and tri-4 (Fig.~\ref{Fig.2}-(a-d)) with each other, 
one can find that the peaks near the Fermi level become stronger, with increasing the defect size. 
This is rationalized: the large defect introduces more carbon atoms at the edges which 
contribute to these peaks. Moreover, as the defect size increases, the interaction of the carbon atoms
at the edges with their images become stronger. As a result, such interaction induces another peak 
under the Fermi level.
Examination the configurations reveals that the C-C bonds at the edges are shortened to be 1.404$\pm$0.04 $\AA$, and 
the C-C bonds between the first shell and the second shell near the edges are enlarged to be 1.433$\pm$0.04 $\AA$. 
Such distortion becomes weak in the third shell (C-C bond length is about 1.419 $\AA$), 
indicating that the defect is localized in space. 

To study the hexagonal holes in a graphene sheet, we first remove the carbon
atoms in a hexagon ring, leaving a hexagonal void in the sheet. Then the shells of carbon atoms 
marked with $S1$, $S2$, and $S3$ in Fig.~\ref{Fig.1}-(b) are removed one by one. 
We name these systems as hex-1, hex-2, hex-3 and hex-4 respectively.
The calculated DOS for these four systems are shown in Fig.~\ref{Fig.2}-(e-h), in which the DOS of the perfect graphene 
is also plotted in Fig.~\ref{Fig.2}-(h) for comparison. Different from that of triangular arrangements, 
the majority spin states and the minority spin states are degenerate for the graphene sheet with hexagonal holes.
Such difference in spin polarization can be understood with Lieb's theorem \cite{Lieb}. 
As shown by Lieb in graphene with a bipartite lattice in the nearest neighbor approximation, the angle 
between the zigzag edges of triangular arrangements is $60^\circ$, and the system is spin splitting. In contrast, 
the angle between the zigzag edges of hexagonal arrangements is $120^\circ$, 
and the system is spin degenerate \cite{Lieb, Furst}.  

As shown in Fig.~\ref{Fig.2}-(e), the DOS of hex-1 change little with respect to that of the perfect graphene, 
which implies that absence of a hexagonal ring in graphene sheet perturbs the electronic structures of the system slightly.
 In the case of hex-2, a weak peak at about $E_{f}+1.2$ eV appears (Fig.~\ref{Fig.2}-(f)).
The local density of state analysis reveals that this peak is contributed from the edge atoms. 
With increasing the defect size, the interaction between the 
defect edges becomes stronger, and the electronic structures of 
the systems change largely, especially near the Fermi level. This is clearly displayed in Fig.~\ref{Fig.2}-(g-h). 
The calculated local density of states reveal that the states near the Fermi level dominantly come from 
the edge atoms of the defect.
After checking the geometries of the systems, we find that for hex-1, the fluctuation of the C-C bonds, 
either shortened to be 1.393 $\AA$ or enlarged to be 1.442 $\AA$, happens within the first shell around the edge. 
With increasing the size of the defect, such distortion of the geometry becomes heavier. 
For the case of hex-4, the structural distortion almost extends to the whole system, 
which eventually gives rise to a small band gap. This is consistent with the previous works \cite{Pedersen-1,Vanevic}.

It is noted that for both triangular and hexagonal arrangements mentioned above, each defect has 
zigzag edges. For comparison, the cases with armchair edges are taken into account. 
A triangular hole with armchair edges, whose size is similar to that of tri-2, is concerned as a typical example. 
Different from that of tri-2, the majority spin and the minority spin states are degenerate. The energy gap is 
about 0.10 eV, and the calculated DOS is like to that of hex-2 somewhat. Moreover, we also investigate the electronic 
structures of a typical hexagonal array with armchair edges, the defect size and the feature of 
the DOS are both similar to that of hex-3, with a gap of about 0.15 eV.

In addition to the triangular and hexagonal arrays, we also arrange the rectangular holes in the graphene sheet. 
A crossed GNR is chosen as supercell, which is shown in Fig.~\ref{Fig.1}-(c), and 
the network of GNRs is shown in Fig.~\ref{Fig.1}-(d).
Following the previous notation \cite{Wakabayashi, Abanin, Louie}, the width of a-GNR is notated with $N_{a}$ 
and that of z-GNR with $N_{z}$, as marked in Fig.~\ref{Fig.1}-(c). 
For convenience, we classify the graphene containing the rectangular holes with using two indexes of $N_{a}$ and $N_{z}$, 
notated as ($N_{a}$, $N_{z}$). For such structural patterns, we consider two series of systems: $N_{a}$ varies from 4 to 12, 
with keeping $N_{z}$ to be 4; and $N_{z}$ varies from 5 to 8, with keeping $N_{a}$ to be 6. 
By calculations, we find that all the concerned systems exhibit semiconducting behaviors.
Moreover, the band gap depends evidently on the width of a-GNR, but weakly on that of z-GNR. 
The band gaps for the concerned cases with varying $N_{a}$ are listed in Table I. From the Table I, one can find 
that the gaps for (6,4), (9,4) and (12,4) are close and much larger, while the gaps for (5,4), (8,4) 
and (11,4) are close and smaller. With increasing the width of a-GNR, 
the gap variation trend seems follow the three-family behavior reported in the infinite a-GNR \cite{Louie}. 
In contrast, for the cases with varying $N_{z}$, the band gaps change slightly. The band gaps are 0.22, 
0.31, 0.09 and 0.14 eV for the cases of (6,5), (6,6), (6,7) and (6,8) respectively.

\begin{table}
  \centering
  \caption{The band gap(in eV) for each case of rectangular arrays of ($N_{a}$,$N_{z}$). $N_{a}$ and $N_{z}$
are the width of a-GNR and z-GNR respectively, which are described in the text.}
                                                                                                                             
\begin{tabular}{|c|c|c|c|c|c|c|c|c|c|}\hline
 ($N_{a},N_{z})$ & (4,4)  & (5,4)  & (6,4)  & (7,4)  & (8,4)  & (9,4)  & (10,4)  & (11,4)  & (12,4)  \\\hline
      band gap  &0.48&0.14&0.68&0.29&0.18&0.75&0.25&0.10&0.71  \\\hline
\end{tabular}
\end{table}

For the rectangular cases, the majority spin and the minority spin states are degenerate. Fig.~\ref{Fig.3}-(a-c) 
show the DOS of a typical three-family system of (7,4),(8,4) and (9,4) respectively. 
Evidently, the band gap is about 0.29 eV for $N_{a}$=7, and it is narrowed to be 0.18 eV for $N_{a}$=8, 
then broadened to be about 0.75 eV for $N_{a}$=9. The similar variation trend is found for the other 
two families. 
In addition, the electronic structures near the Fermi level 
exhibit different features for each case in this family. Typically, in the case of (7,4) (Fig.~\ref{Fig.3}-(a)), 
two peaks emerge at about $E_{f}\pm0.6$ eV, 
which are mainly contributed from the edge atoms of the z-GNRs (z-edge);   
The peaks at about $E_{f}\pm0.3$ eV are ascribed to the atoms which bond to hydrogen 
at the crossing site of a-GNRs and z-GNRs (corner). Moreover, the edge atoms of the a-GNRs (a-edge) 
contribute a little to these peaks. 
 For the case (8,4), the peaks at about $E_{f}\pm0.6$ eV remain, 
while the peaks at about $E_{f}\pm0.3$ eV become flat, as shown in Fig.~\ref{Fig.3}-(b). The 
atoms at the z-edge and a-edge, as well as at the corner all make contribution to these peaks, 
of which the former is dominant. As for the case of (9,4) (Fig.~\ref{Fig.3}-(c)), 
the concerned peaks move to $E_{f}\pm0.5$ eV and $E_{f}\pm0.9$ eV respectively, 
broadening the gap to be 0.75 eV. The former are contributed from the corner atoms, and the latter from 
the atoms at the z-edges. While those at a-edges contribute little to these peaks. 
  Fig.~\ref{Fig.3}-(d-f) shows the DOS for the cases of (6,6),(6,7) and (6,8) respectively.
Similar electronic behaviors near the Fermi level are found in 
these systems: two flat peaks at about $E_{f}\pm0.25$ eV and two peaks at about $E_{f}\pm0.8$ eV. 
The former is mainly contributed from the atoms at the z-edges, and the latter is 
attributed to those at the corner and the a-edge. 

Overall, for the rectangular arrangements, the peaks in DOS near the Fermi level are 
dominantly contributed from the atoms at the edges including a-edges, z-edges and the corners. 
This is essentially ascribed to the large structural distortion near the defect 
edges: the C-C bonds within two neighbouring are 
enlarged (shortened) by about 0.03 $\AA$ (0.05 $\AA$). Moreover, the width of the a-GNR has significant influence on 
the electronic structures of the system, whereas the width of z-GNRs has little. This can be well understood 
from the DOS analysis: the states near the Fermi level dominantly come from the contribution of 
the carbon atoms at the z-edges, but the number of such atoms decrease with increasing the width of a-GNR. 
Therefore, the width of the a-GNR have a crucial role on the electronic structures, 
and this agrees well with that of infinite GNRs, in which the energy gap of a a-GNR significantly dependents on 
its width\cite{Louie}.   

  To better understand the nature of the states in the vicinity of the Fermi level, 
we further explore the charge density for the concerned systems. Some typical cases of the triangular and the hexagonal 
arrangements are given in Fig.~\ref{Fig.4}, 
where the highest occupied molecular orbital (HOMO), the lowest unoccupied molecular orbital (LUMO) 
and the level under the HOMO (HOMO-1) at $\Gamma$ point are plotted. 
For the graphene with zigzag-edged triangular arrangements, the majority spin states for the HOMO-1 and the HOMO 
are of clear localized nature. The former is mainly contributed from the atoms at the edge 
along $y$ axis, and the latter is from those at the other edges. In contrast, the 
minority spin states for the HOMO-1 and the HOMO are delocalized. The majority spin state and the 
minority spin state for the LUMO are mostly confined in three edges. 
This is obviously shown in Fig.~\ref{Fig.4}(a-b) (tri-2 is a representative). 
For the triangular arrangements with armchair edges, the HOMO-1, the HOMO and the LUMO 
states are all of extensive nature, especially for the HOMO and the LUMO states, as shown in Fig.~\ref{Fig.4}(d).
These states, with features of local $\pi$-bond, are contributed from the atoms in the strips along $x$ axis
which contain the defect arrays or not. 

For the graphene containing the hexagonal arrangements with either zigzag edges or armchair edges, 
these concerned orbitals are also localized in a small region near the defect edges: 
the HOMO-1 state is from the atoms at the edges along $x$ axis, and the HOMO and the LUMO states 
are from that along $y$ axis. The typical case of hex-3 and the hexagonal holes with armchair edges 
are shown in Fig.~\ref{Fig.4}(c) and (e) respectively.

From above, we find that for the graphene with hexagonal arrangements, zigzag edges and armchair edges 
make little difference on the electronic structures of the systems. 
In contrast, for the cases with triangular arrangements, edge features
influence the electronic structures significantly. This is rational with considering 
the symmetry and Lieb's Theorem \cite{Lieb}. 
For such different patterns, the size of the defects 
plays a critical role in modulating the electronic structures of the systems .

On the other hand, Fig.~\ref{Fig.4} shows some mixture of a triangularly or hexagonally symmetric edges 
for the HOMO-1 and the HOMO states, corresponding to the triangular and the hexagonal arrangements. 
In fact, the case is the same for the LUMO and 
the level above the LUMO (LUMO+1). 
This is different from the previous literatures \cite{Furst, Vanevic}, where triangular 
or hexagonal symmetry exists in the HOMO or LUMO states. 
This difference is ascribed to the symmetry of a system. 
In the previous literatures, the triangular or hexagonal 
holes were arranged in a hexagonal unit cell, which exhibited high symmetry.  
In contrast, in our concerned cases, the triangular or hexagonal holes are arranged in a rectangular unit cell, 
which makes the symmetry of the bipartite lattice low.
For the cases of the triangular arrangements, only a mirror symmetry along $x$ axis exists in each system. 
While for the concerned hexagonal arrangements, there are two mirror 
symmetries along both $x$ axis and $y$ axis. This is displayed evidently in the isosurface of 
the charge density in Fig.~\ref{Fig.4}. Therefore, it can be concluded that, 
in addition to the features and the size of the defect patterns, the symmetry of the patterns 
is another important factor to influence the electronic structures of the systems. 

 For the graphene with rectangular holes, the charge density displays some interesting features. 
As typical examples, the isosurface charge density for the cases of (8,4), (9,4) and (6,8) 
are plotted in Fig.~\ref{Fig.5}(a-c) respectively.
For the case of (8,4), the HOMO-1 orbital is localized in the separated regions which only belong to z-GNRs. 
This indicates that the electrons are confined in the matrix of z-NGRs. 
In contrast, either the HOMO or the LUMO presents a delocalized state with a major portion in a-GNRs. 
Such a nature implies that a-GNRs in the networks act as some special channels for electronic transport. 
Nevertheless, both the HOMO and the LUMO states exhibit 
local $\pi$-bond, of which the former is from two carbon dimer tilted to $x$-axis (or $y$-axis), while the latter 
is from that along $x$-axis (Fig.~\ref{Fig.5}(a)). Differently, for the case of (9,4), the HOMO-1 
orbital is localized in a-GNRs, and the HOMO and the LUMO states are localized in the crossing areas of 
z-GNRs and a-GNRs, as shown in Fig.~\ref{Fig.5}(b). Meanwhile, both the HOMO and the LUMO 
also exhibit local $\pi$-bond, but the former is from two carbon dimer 
along $x$-axis, while the latter is from that tilted to $x$-axis (or $y$-axis).
Fig.~\ref{Fig.5}(c) is the isosurface charge density for the case of (6,8), which exhibits some new features.
the HOMO-1 state having delocalized nature dominantly comes from the a-GNRs, which also acts as a
channel for electronic transport.
In contrast, the HOMO and LUMO states are all contributed from the atoms at the z-edges.

From above, we find that due to the quantum confinement effect, the electronic states near the Fermi level 
exhibit either localized or delocalized features, which is tightly related to the width of GNRs in the networks. 
As a result, tuning the electronic structures as well 
as wave function of electron may be realized by patterning the graphene sheet.

\section {Conclusion}

Triangular, hexagonal and rectangular holes are arranged in a 
graphene sheet respectively. For these patterned graphene, we systematically investigate their electronic structures, 
based on the first principles calculations. We find that defect patterns with different features can 
 modulate the electronic structures of the graphene significantly. For the cases of the triangular holes 
with zigzag edges, the remarkable peaks appear near the Fermi level, with spin splitting, 
and the peak under the Fermi level converts to be two peaks with increasing 
the defect size. In contrast, for the triangular holes with armchair edges and the hexgonal holes, the majority 
and the minority spin states are degenerate. The graphene with hexagonal arrangements give rise to a narrow band gap. 
For the graphene sheet with rectangular holes, they all exhibit semiconducting behaviors.
The band gap significantly depends on the width of a-GNRs, but weakly on that of z-GNRs. 
Moreover, such structural patterns can induce either localized or delocalized states near the Fermi 
level, and the latter gives rise to the special transport channels along the a-GNR or z-GNR in the networks.

\section *{Acknowledgements}   

 This work is supported by the fundamental Research Funds for the Central Universities, the National Basic Research 
Program of China (Grant No.2009CB939901), and the China Postdoctoral Science Foundation with
code number of 20090450813. The author of H.Y. He gratefully acknowledes the support of K.C. Wong Education 
Foundation, Hong Kong.

\newpage

\begin{figure}
\caption{(Color online) The geometries of the supercell for (a) triangular, (b) hexagonal, 
(c) rectangular arrays in graphene. (d) is geometry of the networks of graphene nanoribbons of (c).}\label{Fig.1}
\end{figure}
                                                                                                                             
\begin{figure}
\caption{(Color online)
The density of states (DOS) for the patterned graphene of (a) tri-1, (b) tri-2, (c) tri-3, 
(d) tri-4, (e) hex-1, (f) hex-2, (g) hex-3 and (h) hex-4. The black solid lines and the blue dot lines
stand for total density of states of the patterned graphene and the corresponding perfect graphene respectively. 
The red dash lines refer to local density of states for the atoms at the defect edges. For (a-d), the 
upper and below curves in each figure stand for the majority spin and the minority spin states respectively.
The Fermi level in each figure is shifted to zero.}\label{Fig.2}
\end{figure}

\begin{figure}
\caption{(Color online)
The density of states (DOS) for the patterned graphene of (a) (7,4), (b) (8,4), (c) (9,4),
(d) (6,6), (e) (6,7) and (f) (6,8). The black solid lines and the violet dot lines stand for total density of states 
of the patterned graphene and the corresponding perfect graphene respectively. 
The short red dash lines, the long green dash lines and the blue dash-dot lines refer to local density of states 
for the atoms at the z-edges, in the corner and at the a-edges respectively. 
The Fermi level in each figure is shifted to zero. }\label{Fig.3}
\end{figure}

\begin{figure}
\caption{(Color online)
The isosurface charge density of HOMO-1, HOMO and LUMO for the patterned graphene of (a) the majority spin 
state and (b) the majority spin state of tri-2, (c) hex-3, 
(d) triangular arrangement with armchair edges, and (e) hexagonal arrangement with armchair edges respectively. 
The isosurface value is $\pm$ 0.05 eV/$\AA^{3}$ (distinguished by orange and blue surfaces).}\label{Fig.4}
\end{figure}

\begin{figure}
\caption{(Color online)
The isosurface charge density of HOMO-1, HOMO and LUMO for the patterned graphene of (a) (8,4), (b) (9,4) 
and (c) (6,8) respectively. The isosurface value is $\pm$ 0.05 eV/$\AA^{3}$
(distinguished by orange and blue surfaces).}\label{Fig.5}
\end{figure}

\widetext

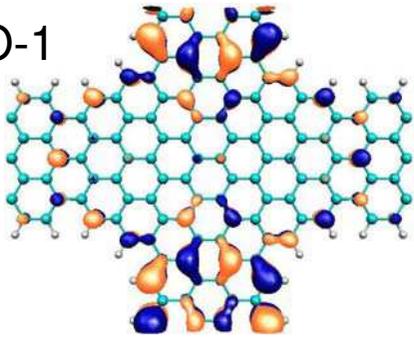
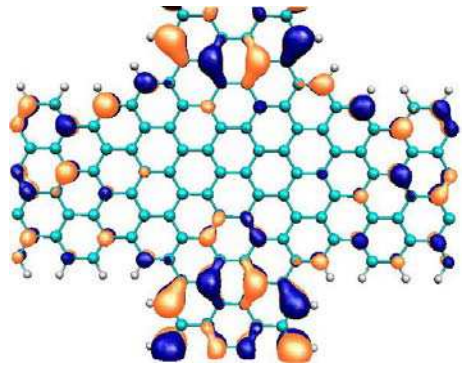
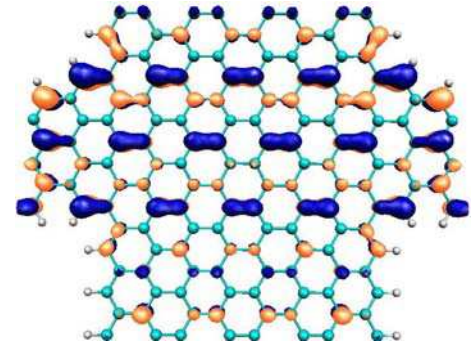
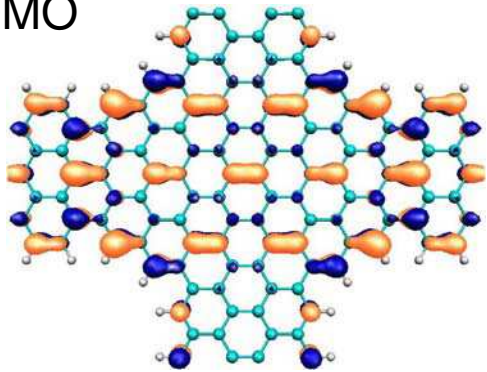
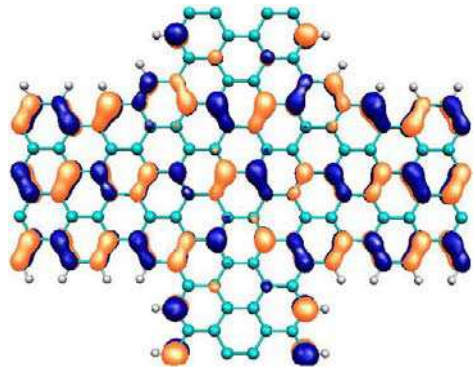
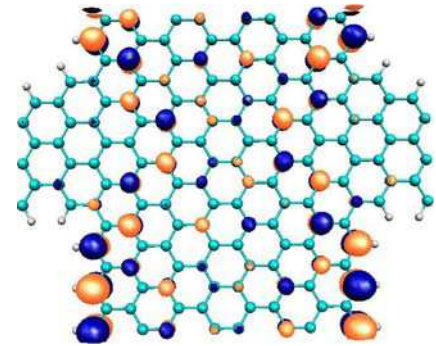
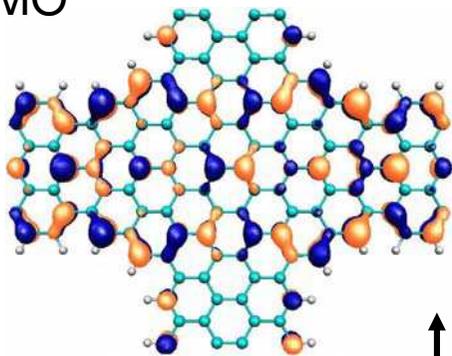
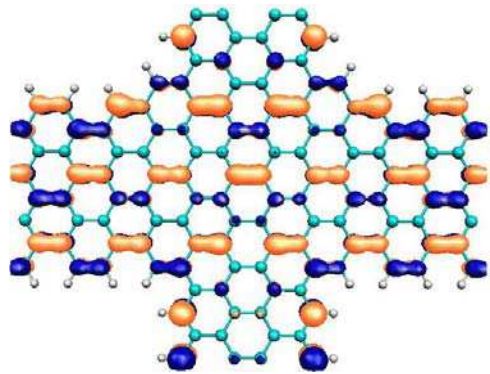
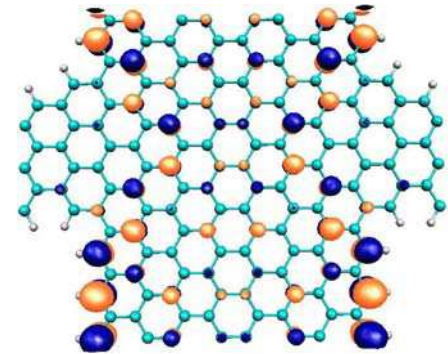
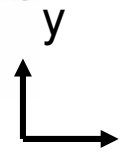
\end{document}